\documentclass{aastex}
\usepackage{emulateapj5}

\shorttitle{{\it Chandra} Observation of Abell~2052}
\shortauthors{Blanton, Sarazin, McNamara, \& Wise}

\slugcomment{to appear in the Astrophysical Journal Letters.}

\begin{document}

\title{{\it Chandra} Observation of the Radio Source / X-ray Gas Interaction
in the Cooling Flow Cluster Abell~2052}

\author{Elizabeth L. Blanton\altaffilmark{1},
Craig L. Sarazin\altaffilmark{1},
Brian R. McNamara\altaffilmark{2},
and Michael W. Wise\altaffilmark{3}}

\altaffiltext{1}{Department of Astronomy, University of Virginia,
P. O. Box 3818, Charlottesville, VA 22903-0818;
eblanton@virginia.edu, sarazin@virginia.edu}

\altaffiltext{2}{Department of Physics \& Astronomy, Ohio University,
Clippinger Labs, Athens, OH 45701;
mcnamara@helios.phy.ohiou.edu}

\altaffiltext{3}{Center for Space Research, Building NE80--6015, 
Massachusetts Institute of Technology, Cambridge, MA 02139--4307;
wise@space.mit.edu}

\begin{abstract}
We present a {\it Chandra} observation of Abell~2052, a cooling flow cluster
with a central cD that hosts the complex radio source 3C~317.
The data reveal ``holes'' in the X-ray emission that are coincident with the 
radio lobes.
The holes are surrounded by bright ``shells'' of X-ray emission.
The data are consistent with the radio source displacing and compressing,
and at the same time being confined by, the X-ray gas.
The compression of the X-ray shells appears to have been relatively
gentle and, at most, slightly transonic.
The pressure in the X-ray gas (the shells and surrounding cooler gas)
is approximately an order of magnitude higher than the minimum pressure 
derived for the radio source, suggesting that an additional source of 
pressure is needed to support the radio plasma.
The compression of the X-ray shells has speeded up the cooling of
the shells, and optical emission line filaments are found coincident with
the brightest regions of the shells.
\end{abstract}

\keywords{
galaxies: clusters: general ---
cooling flows ---
intergalactic medium ---
radio continuum: galaxies ---
X-rays: galaxies: clusters
}

\section{Introduction} \label{sec:intro}

The very central regions of clusters of galaxies are among the most
active physical environments in the Universe.
In a significant fraction of clusters (``cooling flows''),
large amounts of gas ($\sim$$10^2$ $M_\odot$ yr$^{-1}$) are cooling
radiatively from $\sim$$10^8$ to $\sim$$10^7$ K, making these regions
extremely bright X-ray sources
(see Fabian [1994] for a review).
Cooler gas is seen through optical emission lines.
There is always a large (cD) galaxy at the center of these cooling flow
clusters,
and the majority of these galaxies are hosts of radio sources.
It may be that the cooling gas acts
as fuel for the accreting central black hole, thus helping to produce the
radio emission.
Recent {\it Chandra} images have shown that the radio sources and
X-ray gas are interacting;
examples include
Hydra A
(McNamara et al.\ 2000;
David et al.\ 2001)
and Perseus
(Fabian et al.\ 2000).
In both of these cases, there is an anti-coincidence between the radio and
X-ray emission, such that the radio lobes are found in regions where
there are ``holes'' in the X-ray emission.
Depressions in the X-ray emission at the locations of radio emission are
also seen with {\it Chandra} for Abell 2597 (McNamara et al.\ 2001).

Abell~2052 is a moderately rich, cooling flow cluster at a redshift of
$z=0.0348$.
The central cD galaxy, UGC 09799, hosts the complex, powerful radio galaxy
3C~317.
Previous X-ray observations with $Einstein$
(White, Jones, \& Forman 1997),
{\it ROSAT}
(Peres et al.\ 1998),
and {\it ASCA}
(White 2000)
have shown that Abell~2052 has
an average temperature of approximately 3.3 keV, and contains
a cooling flow with a cooling rate of
$\dot{M} \approx 120$ $M_{\odot}$ yr$^{-1}$.
{\it ROSAT} and {\it VLA} images
revealed excess X-ray emission surrounding much of the radio source
(Rizza et al.\ 2000).

In this $Letter$, we discuss a {\it Chandra} observation of Abell~2052
which shows a dramatic interaction between the radio source and X-ray
emission.
We will present a more detailed study of Abell~2052, based on this data, in
a future paper.
We assume $H_{\circ}=50$ km s$^{-1}$ Mpc$^{-1}$ and $q_{\circ}=0.5$ 
(1\arcsec\ = 0.95 kpc at $z=0.0348$).

\section{Observation and Data Reduction} \label{sec:data}

Abell~2052 was observed with {\it Chandra} on 2000 September 3 for a total of
36,754 seconds.
The observation was taken so that the center of the cluster would fall near
the aimpoint of the ACIS-S3 CCD.
Here, we analyze data from the S3 chip only.
The events were telemetered in Faint 
mode, the data were collected with frame times of 3.2 seconds, and the CCD
temperature was -120 C.
Only events with {\it ASCA} grades of 0,2,3,4, and 6 were included.  
Standard bad pixels and columns were removed.
The data were searched for background flares and none were found.
A small period of bad aspect was found and removed, leaving a total exposure
of 36,622 seconds.

\section{X-ray Structure} \label{sec:xray}

Figure~\ref{fig:smooth} shows an adaptively smoothed image of the central
4\farcm2$\times$4\farcm2 region of the cluster.
The smoothed image has a minimum S/N of 3 per smoothing beam and was
corrected for exposure and background, the latter using the blank sky
background fields from Markevitch (2000).
The inner part of the cluster shows the high surface brightness characteristic
of a cooling flow.
At small radii, a bright ring of emission is apparent.
In addition, there is a bar of enhanced X-ray emission running 
approximately East-West through the center.
The brightest parts of the ring and bar appear to form two shells to
the north and south, and the bar may be just the intersection of the two
shells.
These bright shells surround two holes of lowered X-ray brightness.
There is also a spur of emission to the NW of the center which protrudes
into the northern hole.
A point source associated with the center of the cD galaxy and the
central AGN is located above the center of the bar.
Several other point sources are also evident in
Figure~\ref{fig:smooth}.
All of these features are easily seen in the raw image prior to any
smoothing, background subtraction, or exposure corrections.

\centerline{\null}
\vskip3.0truein
\includegraphics{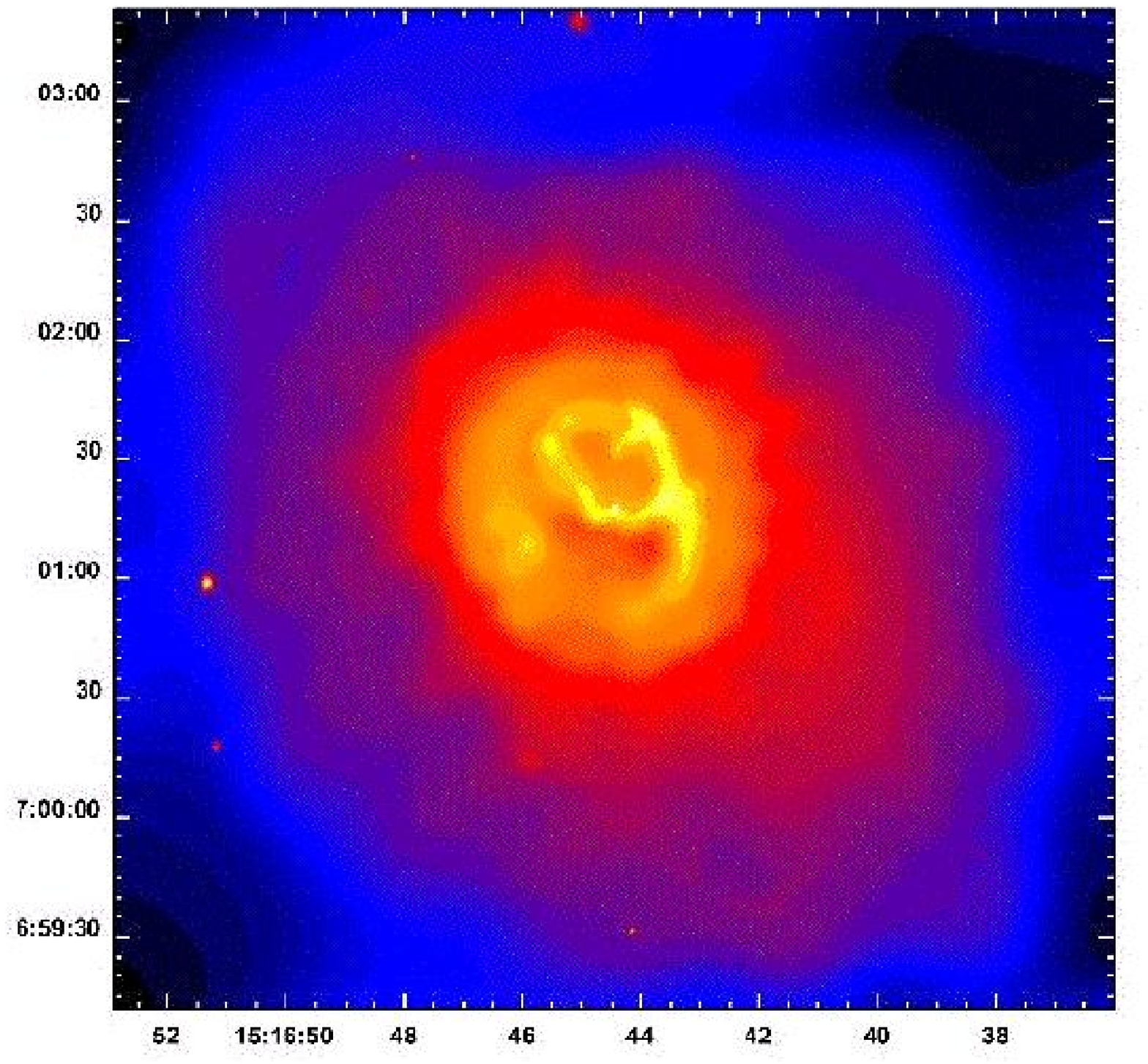}
\figcaption{An adaptively smoothed image of the 4\farcm2$\times$4\farcm2
region surrounding the center of Abell~2052.
The image has been corrected for background and exposure.
The color scale is logarithmic and ranges from
$2.5 \times 10^{-6}$ to $1.1 \times 10^{-2}$ ct pix$^{-1}$ s$^{-1}$.
\label{fig:smooth}}
\vskip0.2truein

Figure~\ref{fig:smooth} gives the impression that
the two holes in the X-ray image are relatively empty regions surrounded
by dense shells of gas, and that the shells are limb-brightened.
To test this, we determined the surface brightness in the centers
of each of the holes.
 From the deprojection of the X-ray surface brightness discussed below,
we estimated the X-ray brightness expected if the holes were indeed
empty, were centered at the same distance as the AGN,
and all of the X-rays were due to projection.
This gave a predicted central surface brightness of
1.5 ct s$^{-1}$ arcmin$^{-2}$, whereas the observed values are
1.5 (1.3) ct s$^{-1}$ arcmin$^{-2}$ for the northern (southern) hole.
As a further test, we determined the mass for the southern shell, assuming it
is spherical and taking the density from deprojection.
We compared this mass to the total mass predicted to be within the
volume of the shell by extrapolating the deprojected density distribution
outside of the disturbed region into these radii.
For the southern shell, the observed mass is $6 \times 10^{10}$ $M_\odot$,
whereas the predicted mass is $(9 \pm 5 )  \times 10^{10}$ $M_\odot$
depending on how the density distribution is extrapolated.
Thus, these numbers are consistent with the idea that the holes are
devoid of X-ray emitting gas, and that the missing gas was pushed
out of the holes and compressed into the shells.

To understand and model the structure in the inner regions of the cluster,
we extracted the X-ray surface brightness
(Fig.~\ref{fig:pressure}a),
and spectra in twenty
circular annuli centered on the central point source,
with radii ranging from 4\farcs7 to 226\arcsec.
Each spectrum typically contained several thousand source counts and
was fitted with a single temperature MEKAL model with the
absorption fixed to the Galactic value
($N_H = 2.85 \times 10^{20}$ cm$^{-2}$;
Dickey \& Lockman 1990).
Background was taken from the blank sky fields
(Markevitch 2000).
The fits reveal that the gas has cooled to a temperature of
$kT\approx1.4$ keV at the center compared to the value of approximately
3.4 keV found for the outer annuli
(Fig.~\ref{fig:pressure}c).

\centerline{\null}
\vskip3.1truein
\includegraphics{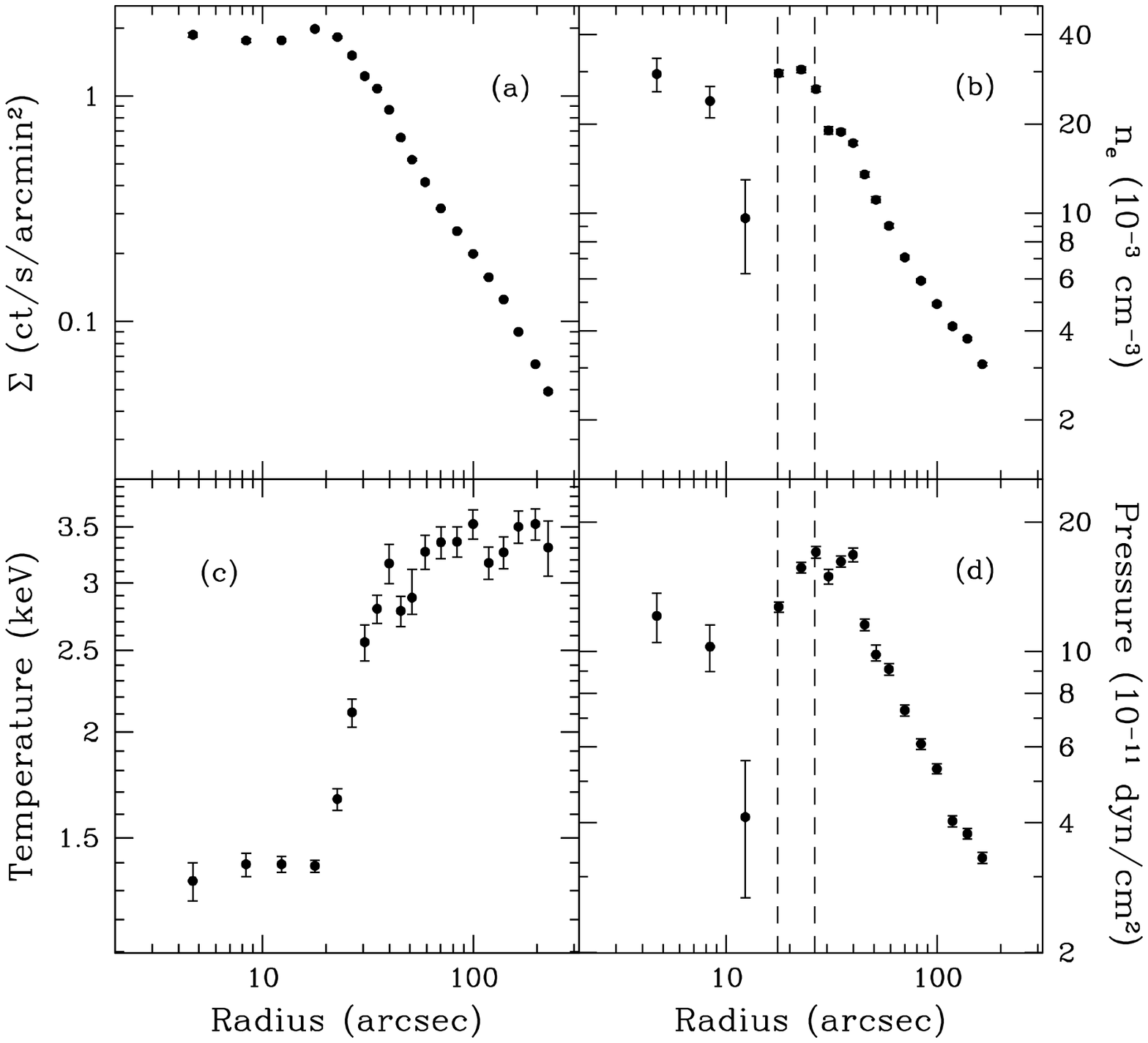}
\figcaption{Surface brightness (a), electron density (b), temperature (c),
and pressure (d), as a function of radius.  The vertical dashed lines mark
the mean inner and outer radii of the bright X-ray ring.
\label{fig:pressure}}
\vskip0.2truein

The surface brightness values for the band 0.3--10 keV
were deprojected to determine the X-ray emissivity and gas
density (Fig.~\ref{fig:pressure}b),
assuming the emissivity is constant in spherical shells.
The density and temperature measurements where used
to determine the radial variation in the pressure in the
gas (Fig.~\ref{fig:pressure}d),
assuming the temperature in the projected spectrum fit is
the temperature at that spherical radius.
Outside of the brightened shell of X-ray emission, the pressure exhibits a
smooth, monotonic decrease with increasing radius, which is presumably
the result of nearly hydrostatic equilibrium.
There is a nearly constant pressure ($P \approx 1.5 \times 10^{-10}$
dyn cm$^{-2}$) at radii corresponding to the bright
ring of X-ray emission and just outside it.
Interior to the ring, the pressure is lower and shows large oscillations.
This is the result of the complex, non-circularly symmetric
structure (the holes, bar, and spur) within this region, and the
spherically averaged pressure is not meaningful.

In order to better determine the pressure in the brightest regions
of the ring of emission,
we measured the pressure in a pie-annular region of the bright ring
located to the West of the center of the cluster.
The surface brightness was measured for the region and corrected
for background taken just outside of the ring.
We assumed that the emission in this region was part of a spherical
shell of emission.
A spectrum was extracted and the pressure was determined in a manner
similar to that described above.
In this case, the pressure was found to be
$P = 1.43 \times 10^{-10}$ dyn cm$^{-2}$, which is consistent with
the values derived assuming spherical symmetry
(Fig.~\ref{fig:pressure}d).

\newpage

\section{Interaction with the Radio Source} \label{sec:radio}

An overlay of the 6 cm radio contours
(Burns 1990)
onto the adaptively smoothed X-ray image (0.3 -- 10.0 keV) of the inner region
of Abell~2052 is displayed in Figure~\ref{fig:radio}.
There is a core source located at the center of the cD galaxy which is
seen in both radio and X-ray.
Most of the extended radio emission is projected within the X-ray holes
to the north and south.
Almost all of the radio emission is contained within the bright X-ray
shells.
However, there appears to be faint radio emission extending slightly
beyond the shells to the south and north where the shell is faint or
absent.

\centerline{\null}
\vskip3.0truein
\includegraphics{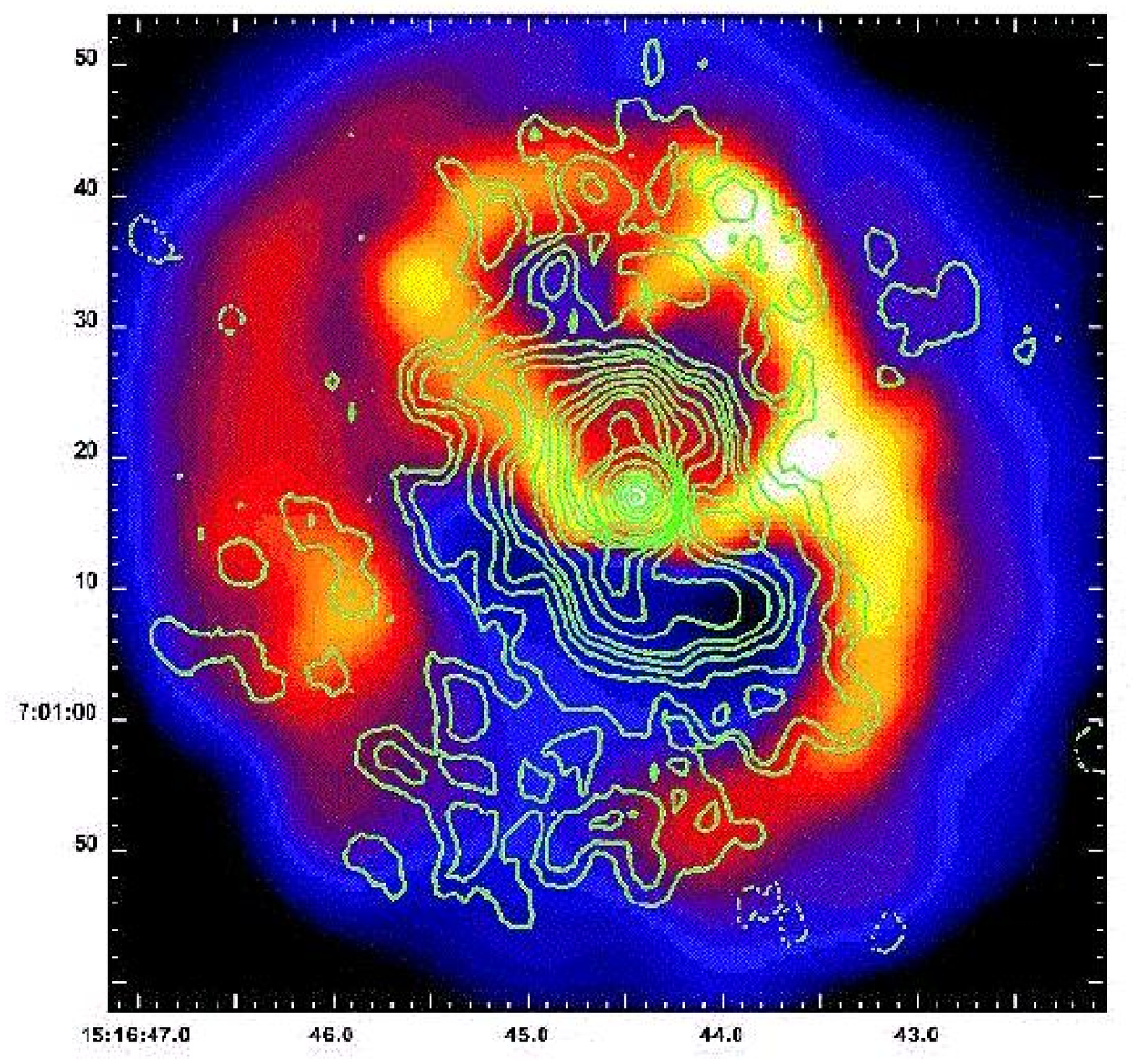}
\figcaption{Radio contours of 3C~317 (Burns 1990) overlaid onto the adaptively smoothed 
{\it Chandra} X-ray image of the central 76\arcsec$\times$76\arcsec\ region of
Abell~2052.
\label{fig:radio}}
\vskip0.2truein

Figure~\ref{fig:radio} suggests that the expansion of the radio source has
displaced the thermal gas which was formerly in the X-ray holes, and
compressed this gas into the bright shells.
As noted above, the X-ray holes do indeed appear to be devoid of X-ray
emitting gas, and the mass of the shells is consistent with the mass
which might have been located in the holes prior to the displacement.
How violent and energetic is the expansion of the radio source?
Heinz, Reynolds, \& Begelman (1998) and
Rizza et al.\ (2000)
suggested the radio source would create cavities in the intracluster gas
by highly supersonic expansion into the gas.
Such a violent expansion would drive strong shocks into the intracluster
gas.
As a result of such strong shocks, the shell of X-ray gas surrounding the
radio-filled cavity would have a higher temperature, pressure, and
specific entropy than the gas just outside the shell.
In fact, this does not appear to be true of the shells in Abell~2052,
although we cannot rule out supersonic expansion by the radio source in
the past when it was much younger.
Our fits to the X-ray spectra indicate that the average temperature of gas
in the shells is
1.1 keV, whereas the temperature of gas just outside the shells is 2.6 keV.
Since the gas in the shells is denser than the gas outside the shells,
the specific entropy is actually much lower than that outside the shells.
Figure~\ref{fig:pressure}d shows that the pressure in the shells is roughly
equal to that just outside the shells, and there is no evidence for a
sharp increase in the pressure which would indicate the presence of a
shock.
The largest increase, which is still roughly consistent with the gradient
in pressure associated with hydrostatic equilibrium,
occurs between radii of approximately 40--45\arcsec.
If we take the pressure increase at this point as an upper limit on
any shock, the Mach number ${\cal M}$ of any shock associated with the
radio source expansion must be ${\cal M} \la 1.2$.
Thus, the expansion of the radio source is subsonic or mildly transonic.
Similar results have been found for the radio source cavities in
Hydra A
(McNamara et al.\ 2000;
David et al.\ 2001)
and Perseus
(Fabian et al.\ 2000).

The relatively slow expansion of the radio source implies that the material
in the radio cavity and the surrounding shell should be nearly in pressure
equilibrium.
Assuming equipartition of energy,
Zhao et al.\ (1993)
calculated the minimum radio pressures for several components of 3C~317
using 6 cm observations.
For their radio ``halo'' component
(the region centered on the core source with dimensions $75\times45$ kpc
which
corresponds to most of the volume of the radio-filled holes),
they determined a minimum pressure of
$P_{\rm min} = 2 \times 10^{-11}$ dyn cm$^{-2}$,
approximately an order of magnitude lower than the {\it Chandra} value
for the X-ray shells.
Even for the radio ``bipolar'' component (a $30 \times 15$ kpc region 
centered on the core), 
they calculated a
value of $P_{\rm min} = 5 \times 10^{-11}$ dyn cm$^{-2}$,
still lower than the value derived from our X-ray observations.

Thus, it seems likely that the radio cavities contain some component with
a larger pressure than given by these equipartition arguments.
This may indicate that one of the basic assumptions of the standard
equipartition arguments
(random pitch angles, magnetic field not parallel to the line of sight, etc.)
is wrong.
The high degree of polarization of the radio source once corrected for
Faraday rotation
(Ge \& Owen 1994)
suggests that the latter assumption might be nearly correct.
There may be a large contribution to the pressure from very low energy
relativistic electrons.
However, the radio spectrum of the large scale radio component is
quite steep ($\alpha \approx -1.9$), so the equipartition estimates
already include a large population of such particles.
The steep spectrum also makes extrapolation to low energies more uncertain.
However, since confined radio sources typically have curved spectra,
the assumption of a power law spectrum should overestimate rather than
underestimate the pressure from the radio source.
The energy and pressure contribution of ions may be greater than
assumed by Zhao et al.\ (1993);
minimum energy arguments would be roughly consistent with the observed X-ray
pressure if the ratio of ions to electrons was $\sim$70, rather than
unity as assumed by Zhao et al.
The magnetic field may be larger than the equipartition value.
Finally, the nonthermal plasma may only contribute a small fraction
of the total energy in the radio-filled cavities (either because
the local nonthermal pressure is low or because the radio plasma
is filamentary and doesn't fill the volume), with the majority of
the pressure coming from very hot, diffuse thermal gas.
We extracted the X-ray spectrum of the southern hole, and
searched for a very hot component in the spectrum.
None was seen, but the observations are not very restrictive.

Cluster cooling flows with central radio sources
are useful bolometers for determining the total energy output of
radio sources, since the energy is apparently confined within the
radio holes by the high pressure intracluster gas rather
than going into adiabatic expansion losses.
Assuming that the pressure within the radio cavities is
$P \approx 1.5 \times 10^{-10}$ dyn cm$^{-2}$  and approximating the
holes as spheres $\sim$20 kpc in diameter, one finds that the total
energy output of the radio source
(including the work done on compressing the intracluster
gas) is about $E_{\rm radio} \approx 1 \times 10^{59}$ ergs.
There is a factor of $\sim$2 uncertainty associated with our ignorance
of the nature of the dominant energy component in the radio source
(higher energy for relativistic particles, lower for magnetic fields).
The assumption of near pressure equilibrium implies that this energy
is comparable to the energy content of the thermal X-ray gas which
filled the holes, but very small compared to the total energy content
of the ICM.
This suggests that individual
radio sources can have dramatic effects locally,
but individual radio outbursts are unlikely to
have large global effects (i.e., on scales much greater than 30 kpc).
On the other hand,
central radio sources are found in most cooling flows, and 
the radio source lifetimes are much shorter than cluster lifetimes;
thus, it is very likely that radio activity is episodic.
The accumulated effect from repeated episodes of radio activity might affect
ICM on larger scales.

The density in the radio cavities is apparently much lower than that in
the ambient gas, and the holes should therefore be buoyant.
They would be expected to rise out of the center of the cluster at
a fraction of the sound speed in the ambient gas, or
$t_{\rm buoy} \sim 2 \times 10^7$ yr.
The buoyant rise time is somewhat longer than the synchrotron lifetime of
the large scale radio source ($t_{\rm syn} \approx 9 \times 10^6$ yr),
which is consistent with the steep radio spectrum
(Zhao et al.\ 1993).
The extended diffuse radio emission at the north and particularly
the south ends of the ring and faintness of the shells in these regions may
indicate that buoyancy and/or related Rayleigh-Taylor instabilities are
acting there.

\section{Relation to Cooler Materials} \label{sec:cool}

We deprojected the surface brightness of the bright western portion
of the X-ray ring, and found a density of $n_{e} = 0.04$ cm$^{-3}$.
Based on the X-ray spectrum of the bright ring, the temperature in
these regions is 1.1 keV.
Using the best-fitting model, we find that the
isobaric cooling time for the gas in the ring is
$t_{\rm cool} \approx 2.6 \times 10^{8}$ yr.
This is much shorter than the probable age of the cluster, but longer than
the upper limit on the age of the radio source $t_{\rm buoy}$.
Thus, it is likely that the gas in the ring is cooling, but 
that most of the compression is due to the expansion of the radio source
rather than cooling.
Note that this compression is likely to have enhanced the cooling;
either adiabatic compression or compression by weak shocks
(${\cal M} < 6.5$ for bremsstrahlung cooling) actually speed up cooling,
despite the associated heating of the gas
(Lufkin, Sarazin, \& White 2000;
Sarazin 2001).

There is also evidence for much cooler material in the same regions.
Contours of H$\alpha$ + [N II] emission from Baum et al.\ (1988)
are plotted over the smoothed X-ray emission in Figure~\ref{fig:halpha}.
The core source is seen at these wavelengths, as well as emission
running roughly East-West through the core and corresponding with 
the X-ray bar.
There is also a component NW of the center, spatially coincident with the 
``spur'' of X-ray emission.
Given the short cooling time in these regions, it seems possible that
this gas at $T \sim 10^4$ K has cooled down from X-ray temperatures,
although this probably occurred before the gas was compressed by the
radio source.
The displacement of the gas from the higher-pressure central regions
outward into lower ambient pressure would lead to adiabatic
expansion, which could also contribute to cooling.
In general, optical line emission is found associated with nearly all
of the very brightest regions of X-ray surface brightness.
This suggests that the H$\alpha$ + [N II] image may be surface-brightness
limited, and that a deeper emission line image would detect emission
from other parts of the X-ray-bright ring.

\centerline{\null}
\vskip3.1truein
\includegraphics{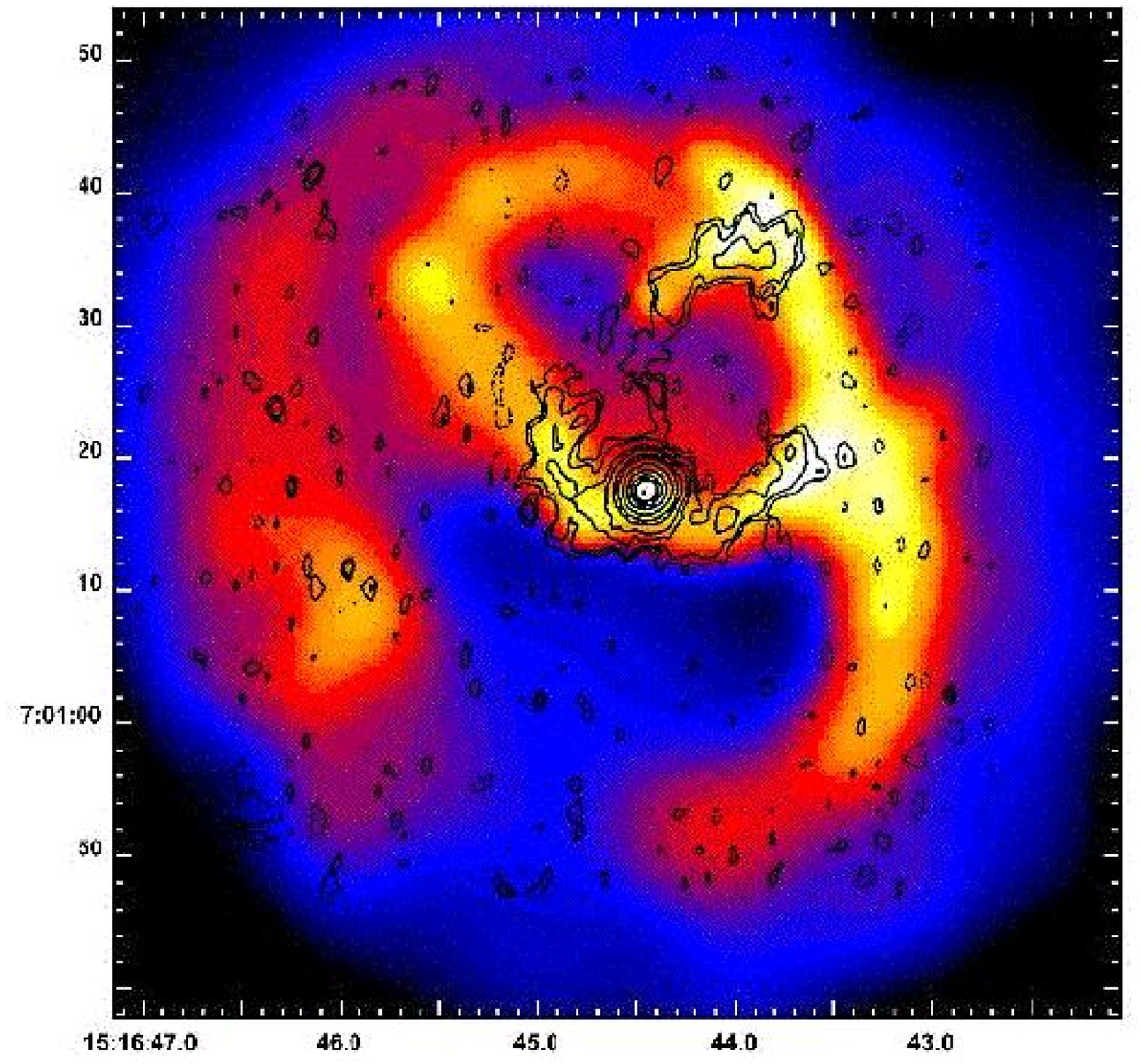}
\figcaption{Overlay of the H$\alpha$ + [N II] contours from Baum et al. (1988) 
onto the {\it Chandra} X-ray image.
The H$\alpha$ + [N II] emission is coincident with the
brightest part of the X-ray ring, consistent with the X-ray spectral
observations that cooling is occurring in these regions.
\label{fig:halpha}}
\vskip0.0truein

\acknowledgements
Support for this work was provided by the National Aeronautics and Space
Administration, primarily through {\it Chandra} Award Number
GO0-1158X,
but also through
GO0-1019X and
GO0-1173X, all
issued by
the {\it Chandra} X-ray Observatory Center, which is operated by the
Smithsonian
Astrophysical Observatory for and on behalf of NASA under contract
NAS8-39073.

\end{document}